\documentstyle[preprint,eqsecnum,aps,prb]{revtex}

\begin{document}
\title{Magnetocapacitance effect in perovskite -superlattice based multiferroics}
\author{M.P.\ Singh, W. Prellier\thanks{%
prellier@ensicaen.fr}, Ch.\ Simon and B.\ Raveau}
\address{Laboratoire CRISMAT, CNRS\ UMR 6508, ENSICAEN,\\
6 Bd du Mar\'{e}chal Juin, F-14050 Caen Cedex, FRANCE.}
\date{\today}
\maketitle

\begin{abstract}
We report the structural and magnetoelectrical properties of La$_{0.7}$Ca$%
_{0.3}$MnO$_3$/BaTiO$_3$ perovskite superlattices grown on (001)-oriented
SrTiO$_3$ by the pulsed laser deposition technique. Magnetic hysteresis
loops together with temperature dependent magnetic properties exhibit
well-defined coercivity and magnetic transition temperature (T$_C$) \symbol{%
126}140 K. $DC$ electrical studies of films show that the magnetoresistance
(MR) is dependent on the BaTiO$_3$ thickness and negative $MR$ as high as
30\% at 100K are observed. The $AC$ electrical studies reveal that the
impedance and capacitance in these films vary with the applied magnetic
field due to the magnetoelectrical coupling in these structures - a key
feature of multiferroics. A negative magnetocapacitance value in the film as
high as $3\%$ per tesla at 1kHz and 100K is demonstrated, opening the route
for designing novel functional materials.
\end{abstract}

\newpage

Superlattices, which are composed of thin layers of two or more different
structural counterparts that stacked in a well-defined sequence, may exhibit
some remarkable properties that do not exist in either of their parent
forms. For example, a (LaFeO$_3$)/(LaCrO$_3$) superlattice stacked on
(111)-SrTiO$_3$ exhibits a ferromagnetic behavior, whereas each parent
material is antiferromagnetic\cite{1}. Similarly, many perovskite
superlattices can exhibit new properties such as high temperature
superconductivity\cite{2}. To grow these superlattices, various thin films
techniques such as molecular beam epitaxy, chemical vapor deposition, pulsed
laser deposition (PLD) technique {\it etc.} have been employed. In
particular, the PLD process is one of the most suitable and frequently used
techniques to grow the superlattice of multi-component perovskite oxides in
a moderate oxygen pressure\cite{3}.

A multiferroic is a material in which ferromagnetism and ferroelectricity
coexist \cite{4}.\ As a consequence, the magnetic domains can be tuned by
the application of an external electric field, and likewise electric domains
are switched by magnetic field. Thus, these materials offer an additional
degree of freedom in designing the various devices, e.g. transducers,
actuators, storage devices, which is unachievable separately in either
ferroelectric or magnetic materials. Hitherto, a very few materials, e.g.
perovskite-type BiFeO$_3$, hexagonal REMnO$_3$ (RE=rare earths), and the
rare-earth molybdates, exist in nature or synthesized in laboratory which
exhibit multiferroism \cite{5,6,7,8,9,10}.\ Note that most of these
compounds display an antiferromagnetic behavior\cite{5,6,7,8,9,10}. These
non-trivial spin-lattice coupling in the multiferroics has been manifested
through various forms, such as linear and bilinear magnetoelectric effects,
polarization change through field-induced phase transition,
magneto-dielectric effect, and dielectric anomalies at magnetic transition
temperatures \cite{5,6,7,8,9,10}. Why and under what circumstances a large
coupling should come about is a major open question, but this problem has
proved difficult to tackle owing to the lack of materials that show such
large coupling. Also the absence of multiferroics with large coupling at
moderate conditions is one of the big hurdles in the realization of
multiferroic devices. Thus, it is essential to design novel multiferroic
materials with essential properties. To synthesize these novel
multiferroics, various efforts have been made by mixing the ferroelectric
and magnetic materials in forms of either composites or multilayers\cite
{11,12} in view of possible applications.

In the present work, we have utilized the versatility of the PLD technique
to create a multilayer structure in a superlattice form composed of a
piezoelectric and ferroelectric, namely BaTiO$_3$ (BTO); and a ferromagnet,
namely La$_{0.7}$Ca$_{0.3}$MnO$_3$ (LCMO). The superlattices were
characterized by the various techniques for their structural, magnetic,
electrical, and magneto-electrical properties, and our results are reported
in this article.

Superlattices of BTO/LCMO were grown on (001)-oriented SrTiO$_3$ (STO) by
the pulsed laser deposition technique, using stoichiometric targets, at 720${%
{}^{\circ }}$C in a flowing 100 mTorr oxygen atmosphere.\ Superlattices with
individual BTO layer thickness of 1 to 25 unit cells (u.c.) by keeping the
LCMO layer thickness as 5 u.c. were realized. The choice of 5\ u.c.\ came
because thin layer of LCMO\ behaves as a ferromagnetic insulator \cite{13}.
The superlattice is composed of 25 repeated units of BTO/LCMO bilayers with
LCMO as the bottom layer.

The samples were characterized by X-ray diffraction (XRD) using Seifert
3000P diffractometer (Cu K$\alpha $ , $\lambda $= 0.15406 nm) and a Philips
X'Pert for the in-plane measurements. Magnetization (M) was measured as a
function of temperature (T) and magnetic field (H) using a superconducting
quantum interference device magnetometer (SQUID). $DC$ electrical
resistivity ($\rho $) of films were measured in four probe configuration. $%
AC $ electrical properties of the films were measured by a lock-in amplifier
(Stanford Research 850) in the frequency range of 1-10$^6$ Hz, where the
sample was held in PPMS system. To measure the electrical properties of the
films in current-perpendicular-to-the-plane (CPP) geometry, a LaNiO$_3$
electrode was fabricated through a shadow mask \cite{14}.

In Fig. 1a, we show the $\Theta -2\Theta $ XRD scan around the (002)
fundamental peak (40${{}^{\circ }}$-52${{}^{\circ }}$ in 2$\Theta $ ) of 5
u.c. LCMO/ 10 u.c. BTO (denoted hereafter as $5/10$) superlattice. The
denoted number $i$ indicates the $i^{th}$ satellite peak. The presence of
higher order satellite peaks adjacent to the main peak, arising from
chemical modulation of multilayer structure, indicates that the films were
indeed coherent heterostructurally grown. The periodic chemical modulation
for the ($5/10$) superlattice, as extracted from $\Theta -2\Theta $ XRD, is $%
\Lambda =5.98$nm which is in agreement with theoretical values ($5.94$ nm)
based on the lattice parameters of each constituent ($0.386$nm for LCMO and $%
0.4006$nm for BTO). The full-width-at-half-maximum (FWHM) of the rocking
curve, recorded around the fundamental ($002$) diffraction peak of the same
superlattice, is very close to the instrumental broadening ($<0.2{{}^{\circ }%
}$), indicating a well crystallinity and a good coherency (see inset of
Fig1a). Further, to examine the in-plane coherence, $\Phi $-scan was
recorded around the $103$ reflection of the cubic unit cell. Different $\Phi
-$scans (not shown here) were recorded by at various tilted angle leading to
a pole figure. Four peaks are clearly observed at $90{{}^{\circ }}$ from
each other, indicating a four-fold symmetry as expected for the perovskite
structures LCMO\ and BTO. This well defined pattern is an evidence for the
in-plane texture of the superlattice. Similar scans recorded on other films
confirm that the superlattices grow epitaxially on STO. The film morphology,
examined by atomic force microscopy, gave a roughness in the range of 3-6A\
(close to one perovskite u.c.) for all films showing that they have a very
smooth surface.

Magnetization vs. applied magnetic field ($M-H$) loop and vs. temperature $%
M(T)$ measurements were performed on all samples and Fig.1b shows an example
in-plane hysteresis loop for a ($5/10$) superlattice, recorded at 10K. The
curve clearly shows a well-defined coercivity confirming the ferromagnetic
nature of the film.\ Furthermore, the temperature-dependance of the
magnetization $M(T)$ recorded under 3000Oe applied magnetic field, shows the
magnetic transition from ferromagnetic to paramagnetic at T$_C$ (Curie
temperature) around 140K.\ This value is lower than the observed bulk LCMO
(250K), but results from both the substrate-induced strain and the thickness
of the layer (5\ u.c.)\cite{15a,15b}. Surprisingly, the Curie temperature of
the superlattices is almost independent of the BTO\ thickness layer (in the
range 135-142K for all films). Despite the similarity in the shape of the
hysteresis loop, the magnetic data reveal some differences.\ For example,
the magnetization of the films is dependent on the thickness of BTO spacer
layer (see inset of Fig.1b). A detailed study shows that the total
magnetization of the films is increasing with increase in the BTO thickness
up to 15 u.c. This is surprising because the LCMO thickness is constant
throughout all the samples. Thus, it is possible to increase the
magnetization only when the LCMO is inducing the magnetization in BTO layer
via magnetoelectric coupling. With further increase in BTO thickness (i.e.
above 15 u.c.) magnetization is decreasing, which is mostly attributed to
the variation in the strains, as previously reported for superlattices\cite
{15a,15b}

Figure 2a displays the{\it \ }$DC$ resistivity of ($5/5$), ($5/10$) and ($%
5/15$) superlattices which were measured as a function of temperature
without magnetic field.\ The inset of Fig.2a displays the magnetic
field-dependance of the magnetoresistance ($MR$) of a ($5/15$) superlattice
taken at 100K. $MR$ is defined as $MR$ $(\%)=100\times [R(H)-R(0)]/R(0)$,
where $R(H)$ and $R(0)$ are the resistance measured with and without
magnetic field, respectively. Fig.2a clearly reveals that the resistivity is
increasing with BTO thickness, which is consistent with the BTO insulating
behavior. With increasing in the BTO layer thickness above 15 u.c., the
resistance of the samples is indeed getting too large to measure above 100K,
as it should be in BTO.\ This shows that the ''magnetic'' BTO is also more
conducting than that of classical BTO.

In order to understand the coherent spin transport in these films, MR were
measured for different samples. Fig. 3b shows the evolution of MR as a
function of temperature for a ($5/15$) superlattice and inset of Fig.\ 3b
shows the $MR$ for several BTO thickness (from 5 to 25) measured at 100K.
Fig.3b clearly indicates that with increasing temperature, the $MR$ is
increasing and vanishes above the transition temperature which is consistent
with the LCMO property. Further, with increase in the BTO thickness up to 15
u.c. (Inset of Fig.3b), there is enhancement in $MR$. This is surprising
because, the $MR$ in a multilayer with an insulating barrier basically
arises from the tunneling of coherent magnetic carriers and spin
polarization of either side of the magnetic layer, i.e. magnetic layer
thickness and structure, whereas in the present case the number of
interfaces and magnetic layer thickness is constant for all superlattices. 
\cite{16}. However, in principle the $MR$ does not increase with increase of
tunnel barrier thickness\cite{16}. The enhancement in MR may be understood
as follows. The total resistivity ( $\rho _{total})$ of the samples can be
expressed as $\rho _{total}$ = $x$ $\rho _{BTO}$ + ($1-x$) $\rho _{LCMO}$
where '$x$' is the BTO compositional resistivity coefficient; $\rho _{BTO}\
and\ \rho _{LCMO}$ is bulk resistivity of BTO and LCMO, respectively . Since
the thickness of the LCMO as well as the number of interfaces are constant
in all samples, the increase in the enhancement MR will be arising from the
BTO layer. Thus, it is evident that, up to 15 u.c., the BTO layer is
magnetically polarized, which corroborates the ($M-H$) findings. Henceforth,
the observed enhancement in magnetoresistance with the increase of BTO
thickness may be attributed to the possible magnetoelectric coupling in
these superlattices as previously seen for (Pr$_{0.85}$Ca$_{0.15}$MnO$_3$%
)/(Ba$_{0.4}$Sr$_{0.6}$TiO$_3$) superlattices\cite{14}. Furthermore, $MR$
was diminishing with further increase in BTO thickness (i.e. above 15 u.c.),
which is attributed to the loss of spin coherence due to the large tunnel
barrier thickness. Thus, the above results (Fig.1b and Fig.2) clearly
exhibit that the maximum magnetoelectric coupling can be expected in ($5/15$%
) superlattice.\ 

In order to study the magnetoelectric coupling in these films, $AC$
electrical properties were measured.\ Fig.3 shows the impedance ($Z$) vs
frequency at different temperatures under 0T and 5T applied magnetic field,
for a ($5/15$) superlattice. Two regimes are observed.\ Below 150K, the
impedance of the sample decreases with applied magnetic field, whereas above
150K it vanishes and no significant variation is seen with magnetic field at
higher temperatures (see Fig.3d). Further, we have extracted the capacitance
of the film based on an equivalent parallel $RC$ circuit, which has been
chosen based on the Cole-Cole plot (not shown) of the observed impedance
data. The estimated capacitance up to 150K is reported in the Fig 4. The
capacitances vs. frequency curves exhibit the two important features. First,
the capacitance of the sample is constant at low frequency and decreases
with the progressive increase in frequency. In a given device dimension, the
capacitance of the device dependence on the total electrical polarizability
of the materials. There are the four electrical polarizability components,
namely electronic, ionic, dipolar and space charge, and their contribution
strongly depend on the frequencies\cite{17}. At low frequency (\symbol{126}%
1Hz -1MHz), the contribution comes from all these components, whereas above
1kHz frequency it starts decreasing and the dipolar contribution becomes
zero around 100MHz. Thus, the decrease of the capacitance at higher
frequency indicates presence of electric dipoles in the film. Hence, it
indirectly provides an evidence of the presence ferroelectricity in these
films. Second, there is a decrease of the capacitance under applied magnetic
fields. In the previous reports, usually the magnetocapacitance effects have
been shown near the Curie temperature. In the present case it occurs well
below the Curie temperature and is significant large in magnitude. The
magnetocapacitance $MC$ is defined, by analogy to the $MR$ as, $MC(T)$ $%
(\%)=100\times [C(H,T)-C(0,T)]/C(0,T)$, where $C(H,T)$ represents the
capacitance at a magnetic field $H$ and a temperature $T$. The negative $MC$
value observed at 1kHz and 100K was \symbol{126}$17\%$. The detail study
reveals that the $MC$ was less than 1\% at 5K whereas it increases with
increasing the temperature and diminishes above 150K. The $MC$ effect, at
low frequency (1kHz) is interesting since it shows that the superlattice can
be considered as a new compound with novel properties that are not observed
in both the parent compounds.

It clearly evidenced that the artificial structure is behaving as a
multiferroic. Such effect was not reported on artificial superlattices. This
effect is a {\it direct evidence of magnetoelectric coupling in the film},
as alluded above, and as reported in other multiferroics\cite{6,7,8,9,10,12}%
. However, it is worth to note that the magnetocapacitance effects in some
ceramics of the hexagonal rare earth manganites are of the order of 1\% {\it %
per tesla }near transition temperature (\symbol{126}40K), whereas in the
present case it is of the order of 3\% {\it per tesla} at 100K. It should
also be noticed that the $MC$ effect appears only at low frequency.\ This
can be understood if one uses the model already proposed in Ref.\cite{12}
where the BTO\ layer is replaced by a $R$ and $C$ in parallel, the LCMO as a
single $R$.\ In this model, the $MR$ of the BTO layer is in agreement with
the $MC$ at low frequency in the whole material.

To summarize, we have successfully grown high quality (BaTiO$_3$/La$_{0.7}$Ca%
$_{0.3}$MnO$_3$) superlattices on STO by PLD process. Despite the lattice
mismatch between substrate and LCMO (-1.17\%), and BTO (+2.2\%), the films
were grown heteroepitaxial. Magnetoelectrical measurements revealed that the
films have at 100K, a negative magnetoresistance close to $4\%$ {\it per
tesla} and a negative magnetocapacitance effect of the order of $3\%$ {\it %
per tesla} at 1kHz.\ The presence of the coupling between dielectric and
magnetic orders have thusly been demonstrated. This high magnetocapacitance
as well as the large magnetoresistance open a path in designing novel
multiferroic thin films.

We would like to thank Prof. James N. Eckstein, Dr. L.\ M\'{e}chin, Prof.
B.\ Mercey and Ms.\ N. Bellido for helpful discussions.

This work has been carried out in the frame of the European Network of
Excellence ''Functionalized Advanced Materials Engineering of Hybrids and
Ceramics'' FAME (FP6-500159-1) supported by the European Community, and by
Centre National de la Recherche Scientifique.

Figure 1: (a) $\Theta $-2$\Theta $ XRD pattern of a ($5/10$) superlattice.
Inset of Fig.1a shows the ($002$) rocking curve of ($5/10$) superlattice,
(b) $M(H)$ curve of ($5/10$) superlattice measured at 10 K. The field is
applied along the [001] direction. Inset of Fig.1b shows magnetic moment as
a function of BaTiO$_3$ thickness at 10 K (Dots are experimental data point
and solid line is just for guiding eyes). The magnetic moment of a LCMO\ (5
u.c.) is 3.10$^{-3}$ emu/cm$^2$.

Figure 2: (a) $DC$ $\rho (T)$ for different superlattices at zero magnetic
field, (b) $MR(T)$ of ($5/15$) superlattice extracted from 0T and 5T $\rho $
data.\ Inset of Fig.2a shows $MR$ of ($5/15$) superlattice at 100K and Inset
of Fig.2b is the $MR$ measured at 100\ K\ as a function of the BaTiO$_3$\
spacer layer.

Figure 3: Modulus of complex impedance ($Z$) vs frequency with applied
magnetic field at different temperatures of ($5/15$) superlattice. Full dots
: 0T, open dots : 5T.

Figure 4: Capacitance vs frequency with applied magnetic field of a ($5/15$)
superlattice at (a) 10 K, (b) 50 K, (c) 100K and (d) 150 K . Full dots : 0T,
open dots : 5T.

\end{document}